\input harvmac
\input epsf

\def\CZ{{\cal Z}}

\def\CO{{\cal O}}
\def\l{\lambda}

\def\CC{ {\cal C}}

\def\p{\partial}

\def\CG{{\cal G}}

\def\CD{{\cal D}}

\def\R{\relax{\rm I\kern-.18em R}}
\font\cmss=cmss10 \font\cmsss=cmss10 at 7pt
\def\Z{\relax\ifmmode\mathchoice
{\hbox{\cmss Z\kern-.4em Z}}{\hbox{\cmss Z\kern-.4em Z}}
{\lower.9pt\hbox{\cmsss Z\kern-.4em Z}}
{\lower1.2pt\hbox{\cmsss Z\kern-.4em Z}}\else{\cmss Z\kern-.4em Z}\fi}
\def\pl{{\it Phys. Lett.\ }}
\def\prl{{\it Phys. Rev. Lett.\ }}
\def\mpl{{\it Mod. Phys.   Lett.\  }}
\def\np{{\it Nucl. Phys.\ }}
\def\pr{{\it Phys.Rev.}}
\def\cmp{{\it Comm. Math. Phys.\ }}
\def\CP{{\cal P}}
\def\r{{\rm Re}}
\def\i{{\rm Im}}

\def\hf{{1\over 2}}

  \def\az{{_{(1)}}}    
  
\def\za{{_{(2)}} } 
\def\aza{{_{(a)}}  }
\def\zaz{{_{(b)}}  }
\def\f{\Phi}
\def\barint{-\hskip -11pt\int}

\lref\GFZ{P. Di Francesco,
P. Ginsparg and J. Zinn-Justin, \pr 254 (1995) 1-133.} \lref\mat{V. 
Kazakov, Phys. Lett. 150B (1985) 282;
F. David, Nucl. Phys. B 257 (1985) 45; D. Boulatov, V. Kazakov, I. 
Kostov, 
and A.A. Migdal, Nucl. Phys. B
275 [FS 17] (1986) 641; 
J. Ambjorn, B. Durhuus, and J. Fr\"ohlich, Nucl. Phys. B 257 (1985) 
43}
\lref\loopeqs{A. A. Migdal, {\it Phys. Rep.} 102 (1983) 199, F. 
David, 
\mpl A5 (1990) 1019.}
\lref\Moor{G. Moore, {\it Matrix models of 2D gravity and 
isomonodromic 
deformations},  in the  
Proc. of the Cargese meeting {\it Random Surfaces and  Quantum 
Gravity}, 
Plenum Press, 1991. }
 \lref\hk{S. Higuchi and I. Kostov,  \pl B 357 (1995) 62.}
  \lref\bkdsgm{E. Br\'ezin and V. Kazakov, \pl B 236 (1990) 144; M. 
Douglas and S. Shenker, \np B 335
(1990) 635;  D. Gross and A.A. Migdal, \prl 64 (1990) 127; \np B 340 
(1990) 333.}
    \lref\kdv{M. Douglas, \pl 238 B (1990) 176; T. Banks, M. Douglas, 
N.
Seiberg and S. Shenker, \pl 238 B (1990) 279. }
 \lref\Inonr{I. Kostov, \pl  B 266 (1991) 317.}
\lref\adem{I. Kostov, \pl  B 297 (1992)74.}
  \lref\Icar{I. Kostov, ``Strings embedded in Dinkin diagrams, in 
the  
Proc.
of the Cargese meeting {\it Random Surfaces, Quantum Gravity and 
Strings},
Saclay Preprint SPhT/90-133.}
\lref\bdks{J. Barb\'on. K. Demeterfi, I. Klebanov amd C. Schmidhuber, 
\np 
B 440 (1995) 189.}
  \lref\Iade{I. Kostov, \np B 326, (1989) 583.}
\lref\ooo{I. Kostov, \pl\   238  B (1990) 181.}
 \lref\mss{G.Moore,
N. Seiberg and M. Staudacher, \np B 362 (1991) 665; G. Moore and N. 
Seiberg, {\it Int. J.  Mod. Phys.} A 7 (1992) 2601.} 
 \lref\ksks{I. Kostov and M. Staudacher,  \pl B 305 (1993) 43.}
\lref\wi{E. Witten, {\it Surv. in Diff. Geom. } 1 (1991) 243.}
   \lref\Idis{I.K. Kostov,  \np B 376 (1992) 539.}
\lref\Iope{I. Kostov, \pl B 344 (1995) 135.}
  \lref\polch{J. Polchinski \np B 362 (1991) 125.}  
 \lref\dvv{M. Fukuma, H. Kawai and R. Nakayama, {\it Int. J. Mod. 
Phys.} A 
6 (1991) 1385;
R. Dijkgraaf, H. Verlinde and E. Verlinde, \np B 348 (1991) 435.}
\lref\kmmm{A. Marshakov, A. Mironov et al., \pl 265 B (1991) 99;
A. Mironov and S. Pakuliak,  {\it Int. Journ. Mod. Phys.} A8 (1993) 
3107;
S. Kharchev, A. Marshakov, A. Mironov, A. Morozov and 
S. Pakuliak, \np B 404 (1993) 717.}
\lref\mirmors{A. Mironov,  A. Morozov and G. Semenoff, hep-th/404005}
\lref\kko{ V. Kazakov and I. Kostov, \np B 386 (1992) 520.}
\lref\onn{I. Kostov, \mpl A 4 (1989) 217, M. Gaudin and I. Kostov, 
\pl 
B220 (1989) 200; I. Kostov and M. Staudacher, \np B  384 (1992) 459;
Zinn-Justin, \np  B 386
(1992) 459; Phys. Lett. B 302 (1993) 396.}
\lref\jimiwa{
 M. Jimbo and T. Miwa, "Solitons and infinite dimensional Lie 
algebras",
{\it RIMS } Vol. 19 (1983) 943-1001.}
 \lref\mehta{
M. L. Mehta, $Random Matrices$  (second edition), Academic Press, New 
York, 1990.}
 \lref\izii{
C. Itzykson and J.-B. Zuber, J. Math. Phys. 21 (1980) 411.}
\lref\acm{J. Ambjorn, L. Chekhov  and Yu. Makeenko, \pl
282 B (1992) 341.}
\lref\ackm{J. Ambjorn, L. Chekhov, C. Kristjansen and Yu. Makeenko, 
\np 
B404 (1993) 127 and erratum \np 449 (1995) 681.}
 \lref\zamsal{ 
P. Fendley and H. Saleur, \np B 388 (1992) 609,
 Al. B. Zamolodchikov, \np B 432 [FS] (1994) 427.}
\lref\ashktel{  Al. B. Zamolodchikov, \np B285 [FS19] (1987) 481}
\lref\Knizh{V. G. Knizhnik, \cmp 112 (1987) 567;
M. Bershadsky and A. Radul, {\it Int. Journ. Mod. Phys.} A. Vol. 2 
(1987) 
165.}
\lref\trw{ 
D. Bernard and A. LeClair, \np B426 (1994) 534;
 C. Tracy and H. Widom, Fredholm determinants and
the mKdV/sinh-Gordon hierarchies, solv-int/9506006, S. Kakei, Toda 
Lattice 
Hierarchy and Zamolodchikov conjecture, solv-int/9510006.}
\lref\DS{Drinfeld and V. Sokolov, {\it. Itogi Nauki i Techniki} 24 
(1984) 
81,
{\it Docl. Akad. Nauk. SSSR} 258 (1981) 1.}
\lref\DFK{P. Di Francesco and D. Kutasov, \np B 342 (1990) 589.}
 \lref\KW{V. Kac and M. Wakimoto, Exceptional hierarchies of soliton 
equations, {\it
Proceedings of Symposia in Pure Mathematics}, 49 (1989) 191}
\lref\Moroz{A. Morozov, Matrix Models as Integrable Systems, 
hep-th/9502091}
 \lref\Pzinn{P. Zinn-Justin, 
{\it Universality of correlation functions of hermitean random 
matrices in 
an external field}, cond-mat/9705044}
\lref\chaoticsp{T. Guhr, A. M\''uller-Groeling and H.A. 
Weidenm\''uller, 
{\it Random matrix theories in quantum physics: Common concepts''}, 
 cond-mat/9707301.}
\lref\verb{J.J.M. Verbaarschot, {\it Universal Behavior of Dirac 
Spectra},
hep-th/9710114}
\lref\Damgaard{G. Akemann, 
 P. Damgaard, U. Magnea and S. Nishigaki,{\it Universality of
 random matrices in the microscopic limit and the Dirac spectrum}
 Nucl. Phys. {\bf B 487[FS]} (1997) 
 721 , hep-th/9609174.}
 \lref\bhz{E. Br\'ezin, S. Hikami and A. Zee,
Nucl. Phys. {\bf B464} (1996) 411.}
\lref\bohigas{O. Bohigas, M. Giannoni, Lecture notes in Physics
 209 (1984) 1; O. Bohigas, M. Giannoni and C. Schmit, Phys. Rev. Lett.
 52 (1984) 1.}
\lref\bipz{E. Br\'ezin, C. Itzykson, G. Parisi and J.-B. Zuber, 
\cmp 59 (1978) 35} 
\lref\Irazl{I. Kostov, {\it ilinear Functional Equations in 2D 
Quantum 
Gravity}, talk delivered at the Workshop on "New Trends in Quantum 
Field 
Theory", 28 August - 1September 1995, Razlog, Bulgaria,  
hep-th/9602117}
\lref\sdalley{ S. Dalley, hep-th-9309068.}
\lref\LPastur{A. Boutet de Monvel, L.Pastur, M.Shcherbina,
 On the Statistical Mechanics Approach to the Random Matrix Theory:
the Integrated Density of States,
 J.Stat.Phys. 79 (1995) 585-611}
\lref\AAk{J. Ambjorn, G. Akemann,  New universal spectral correlators,
cond-mat/9606129}
\lref\KSW{V.A.~Kazakov, M.~Staudacher and T.~Wynter,
\cmp 177 (1996) 451; \cmp 179 (1996) 235;
\np B 471 (1996) 309.}
\lref\KKN{V. Kazakov, I. Kostov, N. Nekrasov, hep-th/9810035} 
\lref\BDeo{E. Brezin and N. Deo, "Smoothed correlators ...", 
cond-mat/9805096}
\lref\krichever{I. Krichever, Comm. Math. Phys. 143 (1992) 415; Comm. 
Pure. and Appl. Math.
47 (1994) 437.}
\lref\takasaki{K. Takasaki and T. Takebe, Rev. Math. Phys. 7 (1995) 
743.}
\lref\thoft{G. 't Hooft, {\it Counting planar diagrams with various 
restrictions}, hep-th/9808113
}
 \lref\morozov{A. Marshakov at al., Phys. Lett.
265B (1991) 99.}
\lref\Vir{ J. Ambjorn, Jurkiewicz and Yu. Makeenko, \pl B () ; 
Itoyama and Y. Matsuo, \pl B (  )   ;  A. Gerasimov et al, \np B  357 
(1991) 565;
A. Mironov and A. Morozov, \pl B 252 (1990) 47.}
\lref\isingi{V.A.Kazakov,
 \pl A 119 (1986) 140;  D.V.Boulatov and V. Kazakov,
\pl B 186 (1987) 379}
 \lref\ACKM{J. Ambjorn, L. Chekhov, 
C. Kristjansen and Yu. Makeenko, \np B 
 404 (1993) 127}.
\lref\AMOS{H. Awata, Y. Matsuo, S. Odake and J. Shiriashi, 
hep-th/9503028.}
\lref\pzj{P. Zinn-Justin, \cmp 194, 631 (1988).}   
\lref\KSW{V.A.~Kazakov, M.~Staudacher and T.~Wynter,
\cmp 177 (1996) 451; \cmp 179 (1996) 235;
\np B 471 (1996) 309.}
\lref\CWM{I.~Kostov and M.~Staudacher, hep-th/9611011;
 I.  Kostov, M.   Staudacher and T.  Wynter,
 {\it Comm. Math. Phys.}  (1998).}
\lref\IZk{C. Itzykson and J.-B. Zuber, Int. Journ. Mod. Phys. A, Vol. 
7 
(1992) 5661.}

  \Title{
\vbox{\baselineskip12pt
 \hbox{SPHT/t98/112}
 }
}
{\vbox{\centerline
{Conformal Field Theory  Techniques }
\vskip2pt     
\centerline{in Random Matrix models     }}}

\centerline{Ivan K. Kostov \footnote{$ ^\ast $}{Member of 
CNRS}\footnote{$ 
^\dagger$}{{\tt kostov@spht.saclay.cea.fr}}}
 \centerline{{\it C.E.A. - Saclay, Service de Physique 
Th\'eorique}}
 \centerline{{\it 
  F-91191 Gif-sur-Yvette, France}}

 \vskip .3in

\baselineskip8pt{

In these notes we explain  how the CFT  
  description of  random 
matrix models  can be used to perform actual calculations. 
Our basic example is the hermitian matrix model,  reformulated as a 
conformal invariant theory of
 free  fermions.     We give an explicit operator construction of the
 corresponding collective field theory in terms of a 
bosonic field on a hyperelliptic  Riemann surface, with special 
operators associated with the 
branch points.  The quasiclassical expressions for
 the spectral kernel and the joint eigenvalue probabilities are then
 easily obtained as correlation functions  of current, fermionic and twist 
 operators. The result for the spectral kernel is valid both in 
 macroscopic and microscopic scales. At the end we briefly consider
  generalizations  in different directions.}
\vskip 1cm

\it{ {Based on the talk of the author at the 
  Third Claude Itzykson- Meeting,  
  "Integrable Models and Applications to Statistical Mechanics ", Paris, 
July 27-29, 1998, and at the workshop ``Random matrices and integrable 
systems'', Univ. of Warwick,
November 2-4 1998.}}
 
 \bigskip

\Date{July 1999 }

\baselineskip=12pt plus 2pt minus 2pt


\newsec{Introduction}

The random matrix models have various applications in rather different 
domains,  and sometimes  language barriers prevents the flow of
ideas and knowledge from one field to another.
 For example, 
such powerful techniques as the conformal field theory (CFT)
description of the random matrix models    and their
relation with the integrable hierarchies,  
which were developped extensively by  the string 
theorists  in the early 90's,  are  practically unknown to the
mesoscopic  physicists.  
  This lecture is an attempt to explain the uses of the
the  CFT description    in a language, which is acceptable by  
both communities. We   therefore  avoided any 
``physical'' interpretation and concentrated on  the method as such. 
The only thing the  
reader is supposed to  
know  are the basics of the two-dimensional conformal field theory. 
 
  \smallskip
  
 \centerline{$\diamond\diamond\diamond$}

 \smallskip

  The statistical ensembles of random matrices of
 large size  (matrix 
models) have been  introduced in 1951 by Wigner  in order 
 to   analyze the spectral 
properties of complicated systems with chaotic behavior \ref\Wign{E. Wigner,
{\it Proc. Cambridge Philos. Soc.} 47 (1951) 790.}.
In this approach the  Hamiltonian of a  chaotic system is considered as a 
large matrix with random entries. Consequently, the analytical studies of 
random matrix 
ensembles carried out in the next 25 years (see   the Mehta's book 
\ref\mEHTA{M.L. Mehta, {\it Random Matrices}, 2nd ed. , New York, 
Academic Press, 1991.}) were oriented to the
calculation of the spectral correlation functions or joint eigenvalue 
probabilities. 

If $M$ is an $N\times N$ hermitian random matrix, then the
spectral correlation function $\rho(\l_1,...,\l_n)$ is defined as the  
 probability  density   that $\l_1,...,\l_n$ are  eigenvalues
 of  $M$   
\eqn\spctrldsts{\rho(\l_1,...,\l_n)={(N-n)!\over N!} 
\left\langle \prod_{i=1}^n \delta(M-\l_i)\right\rangle.}
All spectral correlation functions are expressed as determinants of a single
kernel $K(\l, \mu)$ (the spectral kernel). 
 The spectral kernel can be evaluated by the method of orthogonal polynomials
 \mEHTA .  Its  large $N$ asymptotics 
  is characterized by the interposition of a smooth behavior and
  fast oscillations with wavelength $\sim 1/N$ (in a scale where the total
  range of the spectrum is kept finite).  The smooth  large distance 
  behavior depends on the concrete form of the 
  matrix  potential. On the contrary, the   microscopic behavior  
    characterized  by     oscillations 
    depends  only on the symmetry group (the unitary group in the case)
  and  fall into several universality classes. It is the microscopic 
  behavior of the spectral correlations, which  is
   interesting from the point of view of 
  applications to chaotic systems.
  A review of the latest developments in this direction, in particular 
  these related to  the recent monte Carlo data in lattice 
  QCD, can be found in    \verb .
  
   \smallskip
   
  \centerline{$\diamond\diamond\diamond$}
  
   \smallskip
  The  discovery by `t Hooft 
  of the    $1/N$
   expansion \ref\oovern{G. 't Hooft,  \np B72 (1974) 461.}
    gave a  meaning of the smooth part of the spectral correlators
   and 
  opened the possibility of using random  matrix models
  to solve  various  
combinatorial problems, the simplest of which is  the enumeration of 
planar graphs \refs{\bipz , \thoft}. 
  Here we can mention 
 the exact solution of  various statistical  models defined on random 
surfaces \refs{\isingi, \adem, \onn},
 the
matrix formulation of the 2d quantum gravity
 \refs{\mat , \bkdsgm} and some more difficult combinatorial 
problems as the enumeration of the "dually weighted" planar graphs \KSW\ 
and the branched coverings of a compact Riemann surface \CWM .

In this kind of problems the solution is encoded in the 
 $1/N$ expansion of the 
loop correlation functions, which are the  correlation functions
 of the collective   field variable 
  \eqn\www{W(z)=  \tr\Big( {1\over z-M}\Big).}
 In the large $N$ limit the  correlation 
  functions of the resolvent \www\ are meromorphic functions 
  with cuts along the intervals where the spectral density is nonzero.
  The discontinuity along the cuts gives the
  smooth part of the joint eigenvalue probabilities 
  \spctrldsts.  Indeed, we have the evident relation
  \eqn\dispI{\left\langle W(z_{1})\ldots W(z_{n})\right\rangle_{c}
  = \int 
  {d\lambda_{1}\ldots  d\lambda_{n}
  \over (z_{1} - \lambda _{1}) \ldots
  (z_{n} - \lambda _{n}) }\ 
  \rho(\l_1,...,\l_n)
  }
  where $\langle \ \ \ \rangle_{c}$ means connected correlator.

   The first exact results
in the large $N$ limit were obtained by direct
application of the saddle point method \bipz , 
 but  later it was recognized that a  more powerful method is 
provided by the 
 so called loop equations \loopeqs  , whose iterative
solution  allows one to  reconstruct order by order the $1/N$ expansion. 
 The most efficient iterative procedure proved to be the 
  "moment's description" \ackm , which allowed to 
calculate the free energy and loop correlators
for an arbitrary potential up to $1/N^4$ terms.

\centerline{$\diamond\diamond\diamond$}

In the  early 90's,  after 
the publication of  the 
seminal papers \bkdsgm , the two resolution techniques in random matrix models 
 (orthogonal polynomials and loop equations) developed  rapidly  
 and  were recognized as particular cases of well developed 
 mathematical methods.

 The method of orthogonal polynomials was reformulated in terms of  the
 theory of $\tau$ functions of integrable hierarchies  (see, for example, 
 the review article  \GFZ ). On the other hand it 
 was observed that the loop 
equations  generate a representation of (half of) the Virasoro algebra and 
are therefore equivalent to the requirement  that the theory is {\it conformal 
invariant} \refs{\dvv, \kmmm}.   
This allowed to describe  a class of large $N$ matrix models near 
criticality in terms of  the
Hilbert space of   twisted bosonic fields, and apply 
the well developed formalism
of the conformal field theory. Finally, it was observed that the two 
approaches are closely related since the $\tau$ functions associated with
the matrix models can be formulated as fermionic theories with 
conformal invariance \Moroz .  (The approach based on  integrable
hierarchies is however less general since it works only for the
 unitary ensemble of random matrices.)

 \smallskip
 \centerline{$\diamond\diamond\diamond$}
 \smallskip
 
  %
  %
  It happened   that the CFT techniques were
 developed exclusively from the point of view of their
 application to   noncritical
string theories, and therefore are known only to a  relatively narrow circle 
of physicists.   
 The aim of this lecture is to demonstrate the 
technical  and conceptual  advantages of  the conformal field theory 
description 
  in a more general
setting. In particular, we will  reproduce, using  the CFT formalism,  
 the microscopic oscillations    of the
spectral kernel and  correlators.  The CFT formalism allows to 
understand better the origin of the observed universality at mesoscopic 
and microscopic scales.

  %
  %

 We will restrict ourselves to the   simplest example of 
 the hermitian one matrix model with
arbitrary potential and will only briefly  consider 
the generalizations at the end.   
   First we will establish the 
equivalence of the unitary matrix ensemble with a system  of 
two-dimensional free chiral fermions.  Then we will construct  the 
collective field theory, which is obtained from the fermionic system 
by the  two-dimensional bosonization rules.  
  The field  posesses  a nontrivial expectation value  associated 
with  a 
hyperelliptic Riemann surface.  
  The latter is  described, from the point of view of CFT, by a 
collection 
of twist operators
associated with its branch points.
 with interaction concentrated at the branch points.  
We will demonstrate the power of the 
CFT description  
by performing some actual calculations.  
We will obtain the  
     quasiclassical 
expressions for the correlator of the resolvent 
$W(z)$,  the
joint eigenvalue probabilities and the spectral kernel. 
Finally we will briefly mention how the
 formalism  can be generalized to
the case of a chain  of 
coupled random matrices in presence of 
  external matrix fields.

 
\newsec{ The hermitian one matrix  model. Loop equations 
as Virasoro constraints}

The partition function of the 
unitary ensembleis defined as the integral 
 \eqn\hemm{
\CZ_N[V] =  \int dM \ e^{-\Tr V(M)} }
where $dM$ denotes the translational invariant measure in the space 
of  
$N\times N$   hermitian 
matrices $M=\{M_{ij}\}_{i,j=1}^N$.
 We normalize the    measure as
\eqn\normM{ dM = {1\over {\rm Vol} [U(N)] }   \prod_{k=1}^N  \  
{dM_{kk}\over  2\pi } \prod _{k<j}  2 \ d\r (M_{kj}) 
\ d\i (M_{kj} )}
where
${\rm Vol} [U(N)] = \prod_{k=1}^N {(2\pi)^k\over k!}$
 is the volume 
 of the
unitary group.
 We assume that the potential is   
an entire function 
 \eqn\potnTl{V(M) = - \sum_{n=0}^\infty t_n M^n } 
which is the natural choice for the ensemble of hermitian matrices.

The integrand depends on the matrix variable $M$ only through its  
eigenvalues 
$\lambda_1,..., \lambda_N$ and the integral   
\hemm\   is actually reduced to 
\eqn\Coulgaz{\CZ_N[V] = {1\over N!}\ \int   \prod _{i=1}^N    
d\lambda_i \ 
e^{- 
V(\lambda_i)}  \ \prod_{i< j} (\l_i-\l_j) 
^2.}

   The loop equations   represent an infinite set 
of identities satisfied by 
the 
 correlation functions of the collective   field variable 
  \eqn\www{W(z)=\sum_{i=1}^N {1\over z-\l_i} = \tr\Big( {1\over 
z-M}\Big).}
The    derivation goes as  follows.   From the translational 
invariance of
 the integration measure $d\lambda_i$ 
we find 
\eqn\dseon{ 
\Big\langle \sum_{i=1}^N  \Big({\p\over\p\lambda_i}+
2\sum_{j(\ne i)} {1\over \lambda_i-\lambda_j}  
+\sum_{n\ge 0}nt_n\lambda_i^{n-1} \Big){1\over z-\lambda_i} 
\Big\rangle_{N,\{ t\} }  =0 }
where $\langle \ \ \ \rangle_{N,\{ t\}}$ means the average with respect to 
this 
partition function.
  Using the identity
\eqn\sdaa{
\sum_i {1\over (z-\lambda_i)^2}+2\sum _{i\ne j} {1\over 
z-\lambda_i}{1\over \lambda_i-\lambda_j}=
\sum_{i,j} {1\over z-\lambda_i} {1\over z-\lambda_j}}
we  write \dseon\ as
\eqn\dsa{ 
\Big\langle W^2(z)+  
\sum_{i=1}^N {1\over z-\lambda_i} \sum_{n\ge 0}nt_n\lambda_i^{n-1}  
\Big\rangle_{N,\{ t\}} =0.}
 Assume that the point $z$ is very far from the origin.  Then the sum 
in 
the second term can be viewed as the result of  a 
 contour integration 
   along a big  contour $\CC$ circling   all the
eigenvalues $\l_1,...,\l_N$ 
 but not the point $z$:  
\eqn\eqct{  \oint_{\CC}
{dz'\over 2\pi i} {1\over z-z'} \Big\langle
{   T}(z')   
\Big\rangle_{N,\{ t\}} =0 }
where  
\eqn\colf{{  T}(z)=  \hf[\p\Phi(z)]^2, \ \ \ \Phi(z)={1\over \sqrt{2} 
} \sum_{n\ge 0}
t_n z^ n -\sqrt{2}\ \tr \log \Big({1\over z-M}\Big). }
  The insertion of the operator  $\tr M^n =\sum_i \lambda_i^n$ can be 
realized by taking 
 a partial derivative with respect to the coupling $t_n$.
This allows to represent the operator \colf\ as  
 \eqn\bfPhi{{ \bf T}(z)= \hf  [\p{\bf \Phi}(z)]^2, \ \ \  
 {\bf \Phi} (z) ={1\over \sqrt{2}} \sum _{n >0}
t_n z^{n} +  \sqrt{2}N \ln z + 
 \sqrt{2} \sum _{n\ge 0}
{z^{-n}\over n} {\p\over \p t_n} }
and write the loop equations as  linear differential constraints 
  \eqn\viraa{ {\bf L}_n\cdot \CZ_N[t]=0 \ \ \ \ \ \  ( n\ge -1)}
 where
\eqn\lmOm{ 
{\bf L}_n=\sum_{k=0}^{n}   {\partial\over \partial t_k}
 {\partial\over \partial t_{n-k}}+
\sum_{k=0 }^{\infty} k t_k  {\partial\over \partial t_{n+k}}.
   }
   (We used the notation $\CZ_{N}[t]$ instead of  $\CZ_{N}[V]$ 
  since a concrete representation of the  the potential $V(z)$ 
    by the  set  $\{ t_{n}\}_{n=0}^{\infty}$ of 
   coordinates   (coupling constants) 
   is chosen.)
The operators ${\bf L}_n $  satisfies  the Virasoro algebra
  \eqn\viR{[{\bf L}_m,{\bf L}_n]=(m-n){\bf L}_{m+n}}
which implies that the   this  matrix integral 
 realizes a representation of the    conformal 
 group.   In writing the Virasoro constraints we 
introduced the derivative with respect to the trivial variable $t_0$, 
which satisfies
  \eqn\trto{{\p \over \p t_0} \CZ_N [t]= N \CZ_N [t].}

It turns to be  much simpler to use apparatus of the conformal field 
theory instead of solving directly the loop equations \viraa .
 The formal solution of these equations is given in terms of
a chiral Dirac fermion or, through bosonization,  of   a chiral 
bosonic field with Liouville-like interaction, which is
by its definition conformal invariant. Such a construction 
has been proposed  in  \refs{\kmmm , \Moroz}. 
 
In the next sections  we consider the CFT interpretation of the 
matrix 
model
from more pragmatic point of view than in refs.  \refs{\kmmm , 
\Moroz}. 
Our aim is  explore the  large $N$ limit of the operator solution 
of the loop equations through the conformal field theory 
and reproduce the $1/N$ expansion for the free energy and the
correlation functions.
 The bosonic   field can be considered as the 
collective field describing the Dyson gas of eigenvalues.
The $1/N$ expansion can be constructed in terms of a free bosonic 
field 
defined on a Riemann surface 
 determined by the classical background (the spectral
density in the thermodynamical limit).
 The spectral correlations can be determined directly 
from the correlation functions of this bosonic field.

\newsec{Fermionic representation}

 \subsec{ Fock space representation   of the hermitian 
matrix integral  in terms of Dirac fermions}

  The model \hemm\ was first solved by the 
  so called method of orthogonal polynomials
 \refs{\mehta, \izii}. The method is  based on 
the possibility to consider the matrix model as a
system of free fermions. Later the 
  fermionic partition function was identified as
a $\tau$-function of the KP hierarchy  \Moroz .
 (For a review of the theory of tau-functions see \jimiwa .)

There are different fermionic representations
of the matrix integral  depending on the choice 
of the fermionic wave functions.  Below we
will describe the one that is most  natural   
from the point of view of the conformal symmetry.
Let us introduce the   chiral fermions 
\eqn\fermionns{\psi(z) = \left\{ \matrix{ \psi ^\az (z)\cr
 \psi ^\za(z)\cr}\right\}, \quad 
  \psi^{\dag}(z) = \left\{ \psi^{*  \az  } (z),
\psi^{*  \za  }  (z)\right\}.}
Choosing a local
  coordinate $1/z$ at $z=\infty$ we expand
\eqn\pzpo{
        \psi^\aza (z)= \sum_{r\in \Z+ {1\over 2}}\psi_{r}^\aza 
z^{-r- {1\over 2}}, \ \ \ \ 
        \psi^{*  \aza  } (z)= \sum_{r\in \Z+ {1\over 2}}
\psi^{*  \aza  }_{ r} z^{-r- {1\over 2}} .  }
where the fermion modes satisfy  the anticommutation relations
 \eqn\cpmto{
        [ \psi ^\aza
_{ r},\psi^{* \zaz}_{ r'} ]_+= \delta_{a,b} \delta_{r+r',0}. 
}
 The left and right 
vacuum states  of given charge $\vec l = (l^{^{(1)}}, l^{^{(2)}} )$ 
are defined by
\eqn\mnfio{
\eqalign{
        \langle \vec l |     \psi_{r}^{\aza} =\langle \vec l| 
\psi^{*\aza }_{r}&
= 0\ \ \ \
         \ (r<l^\aza)\cr
\psi_{r}^\aza | \vec l \rangle =\psi^{*\aza}_{r}| \vec l\rangle & = 
0\ \ \ \
 \ (r> l^\aza) \cr}
        }
 and the  corresponding normal ordering   is
denoted by $:\quad : $.
The action 
 of the chiral fermions is $S= \int d^2 z
 \psi^{\dag } \bar \p \psi  $
and the energy-momentum tensor is $T= \hf 
[ \p \psi^{\dag }\psi -
\psi^{\dag} \p \psi]$.
 
The fermion bilinears 
\eqn\currennts{\eqalign{& J^\aza (z) = :\psi^{*\aza}(z) 
\psi^\aza(z):    
\quad\ \  ( a = 1,2) \cr
& J_+(z) =
\psi^{*\az} \psi^\za(z),\ \ J_-(z) =
\psi^{*\za} \psi^\az(z)\cr}}
satisfy the $u(2)$ current algebra. 

 The currents $J^\aza(z) = \sum_n J_n^{\aza} z^{-n-1},$
  $J_n\aza= \sum _{r\in \Z+\half} :\psi^{*\aza}_r\psi^\aza _{r+n}: $ 
commute  with the 
 fermionic fields as
 \eqn\cmmr{\eqalign{& [\psi^\aza(z),J_n^\aza]= z^{n} \psi^\aza(z), \ 
\ 
[J_n^\aza,\psi^{*\aza}(z)]=z^n 
  \psi^{*\aza}(z),\cr
  & [\psi_r^\aza, J_n]=\psi_{r+n}^\aza, \ \ 
[J_n^\aza,\psi_r^{*\aza}]=\psi^{*\aza}_{r-n}\cr & 
[J_n^\aza,J_m^\aza]=n\delta_{n+m, 0}.\cr}}
We will also  denote the $su(2)$ current components by
\eqn\utwo{\eqalign{E_+(z)&=J_+(z) , \cr
E_-(z)&=J_-(z) , \cr
 H(z) \ &= \hf[J^\az (z)- J^\za (z) ] \cr}}
 and the $u(1)$ current by
\eqn\uone{ 
\tilde H(z) = \hf[J^\az (z)+ J^\za (z)  ].}
 The $su_{2}$ currents are  expressed in 
terms of the Pauli matrices as
\eqn\curPaul{ \eqalign{
H \ &= :\psi^{* } \sigma_{3} \psi : \ , \cr 
E_{\pm} & = :\psi^{* } \sigma_{\pm} \psi :  \ .\cr}
 }

Now we are ready to write the Fock-space representation
 of the matrix partition function   \hemm \   
in terms of the 
fermionic fields \fermionns . 
 For this purpose we define the "Hamiltonian"
\eqn\nmltn{\eqalign{H[V]\ &=-\oint {dz\over 2\pi i}
V (z)  H(z) \cr &= \hf\sum_{n\ge 0} t_n \sum_{r} \left(: 
\psi^{*\az}_{n-r} 
\psi^{\az}_r -\psi^{*\za}_{n-r} \psi^{\za}_r:\right)\cr}
}
 and the 
"screening operator"
\eqn\glinfty{ Q_+ = \int_{-\infty}^{\infty}  
 d\l\ E_+(\l ).    }

\bigskip

\noindent
{\it Statement:}
 
\bigskip

\noindent
 The scalar product 
\eqn\fockrp{\eqalign{ \CZ_{N}[V] =
\langle N|   e^{H [V]}  \ e^{Q_+} |0  \rangle},}
where $|0  \rangle \equiv |0, 0  \rangle $ and
$\langle N|\equiv \langle N, -N| $, is equal to the   integral 
\Coulgaz .

\bigskip

 \noindent
{ \it   Proof:} 
\bigskip

First we expand $e^{Q_+}$ in fermion multilinears 
and move the operator $e^{H[V]}$ to the right
 using  the  formulas
\eqn\evolU{\eqalign{ e^{H[V]} \psi^\za (z)e^{-H[V]}
 &= e^{-\hf V(z)}  \psi^\za (z),\cr
e^{H[V]} \psi^{*\az} (z)e^{-H[V]}
& = e^{- \hf V (z)}  \psi^{*\az} (z),\cr}
}
until it hits the right vacuum. The result is\foot{Here 
we assume that the potential grows sufficiently fast at 
$\l \to \pm \infty$ and therefore
the  integral is finite if the fermionic operator
is regular at $z=0$. This is indeed the case since all
$\psi_r$ with $r>0$ are annihilated by the right vacuum.
 }
 \eqn\taUu{
 \CZ_{N}[V] =
\langle  N| e^{ Q_+^V  } |0 \rangle,
 \quad Q_+^V  =    \int_{-\infty}^{\infty}  d\l 
   \psi^{*\az}(\l)\psi^\za (\l )\ e^{-V(\l)}
 .}
Then, expanding the exponent in modes
\eqn\lnOm{ Q_+^V= 
  \sum_{r,r'>0} \int d\l  \l^{r+r'-1} 
\psi^{*\az}_{-r}\psi^\za_{-r'}e^{-V(\l)}}
and using  \cpmto \ and \mnfio , we reproduce the original integral.

\bigskip

The representation \fockrp\ of the partition function 
implies that  the latter is a $\tau$-function of the KP hierarchy 
(this 
fact has been established in \morozov)
and therefore satisfies an infinite tower of differential equations, 
the 
first of which is the KP equation.

 \subsec{Relation to the method of orthogonal polynomials} 

The  method of orthogonal polynomials \mehta\ 
 consists, in this setting, 
 in choosing a basis 
diagonalizing the quadratic form \lnOm 
\eqn\dfpn{ 
  \sum _{r\ge {1\over 2} } 
z ^{r-{1\over 2}} 
\psi_{-r}^{* \az}  
 =\sum_{n\ge 0} P_n(z ) b_n^{*\az},  \quad
  \sum _{r\ge {1\over 2} } z^{r-{1\over 2}} 
\psi_{-r}^{\za}  |0\rangle
         =\sum_{n\ge 0} P_n(z) b_n^{\za}  }
where the polynomials $P_n(z)= z^n+...$ satisfy the orthogonal 
relations
\eqn\orTT{\int _{-\infty}^{\infty}  d\l \  P_n(\l) P_m(\l)e^{-V(\l)}= 
h_{n+1}
\delta_{m,n}.}
The operators $b_n^{*\az} $ and $b_n ^{\za}$ are linear combinations 
of   $\psi_{-r}^{* \az}$ and $\psi_{-r}^{\za}$ respectively:
\eqn\bbpsipsi{b_n^{*\az}=\sum _k B_{nk}\psi_{-k-\hf}^{* \az} , \quad
b_n^{\za}=\sum _k B_{nk}\psi_{-k-\hf}^{ \za}}
where the matrix $B$ is triangular and with  with unitary diagonal 
elements
\eqn\Bbb{B_{nn}=1, \quad B_{nk} = 0 \ \ {\rm for } \  k>n}
and therefore its determinant is equal to one, $\det B=1$.

Introducing the set of orthogonal  functions
\eqn\ortbaZ{ f_n(\l) =P_{n-1}(\l)e^{-\hf V(z)}, \ \  n=1,2,...}
such that $ (f_k, f_j) \equiv \int d\l f_k(\l)f_j(\l)  =\delta_{kj} 
h_k$,
we represent the quadratic form  \lnOm \  as
\eqn\gdgn{\eqalign{Q_+^V =   \sum_{m,n=1}^{\infty}
 (f_m,   f_n) \  b^{*\az}_m 
b_n^\za = \sum_{n=1}^{\infty} h_n \  b_n^{*\az}b_n^\za}}
and  the partition function equals
\eqn\ptffF{\CZ_N[V]= {1\over N!}(\det  B)^2 \ {\det}_{_{N\times N}} 
(f_k,f_j) = 
h_1h_2...h_N.}

\subsec{Spectral kernel and joint eigenvalue distributions}

 A complete set of observables 
in the matrix model is given by the 
 joint distribution  probabilities
 for $n$ eigenvalues ($1\le n\le N$) 
\eqn\jevp{\rho(\l_1,...,\l_n) =  {(N-n)!\over N!}
\Big\langle \prod_{k=1}^n 
\delta(\l_k -
 M\Big\rangle .}

The density probabilities \jevp\ are expressed as 
expectation values of fermionic bilinears. Indeed, keeping the only
relevant term in the  expansion   $e^{Q_+}= \ldots + 
Q_{+}^{N}/N!+\ldots$ 
 we  write the partition function as
\eqn\homogn{\CZ_N[V]= 
\int { d \l_1 ... d\l_n \over N!} \ \left\langle N \left| 
e^{H[V]}\prod_{k=1}^N
E_+(\l_k)  \right|0\right\rangle = \int d \l_1 ... d\l_n \ \rho(\l_1, 
..., 
\l_n)}
from where the representation of the probability measure 
$\rho(\l_{1}, \ldots , \l_{N})$ 
as the expectation value of $N$ operators $E_{+}$.
  Integrating with respect to  $N-n$ of the $\l$'s gives
\eqn\jevP{
\rho(\l_1, ... , \l_n) ={(N-n)!\over N!}
\left\langle  \prod_{k=1}^n E_+(\l_k)\right\rangle = {(N-n)!\over N!}
\left\langle  \prod_{k=1}^n \psi^{*\az}(\l_k)\psi^{\za}(\l_k)
\right\rangle }
where by definition $\langle \CO\rangle = {\langle N|
e^{H[V]} \CO e^{Q_+}|0\rangle \over  \langle N|e^{H[V]} 
e^{Q_+}|0\rangle}$.
 In particular, the spectral density is the expectation value 
of the fermionic current
\eqn\dnsf{\rho(\l) =     {1\over N}
\langle   \psi^{*\az}(\l)\psi^{\za}(\l)\rangle.}

Since we have a system of free fermions,  the average 
is equal to the determinant of the two-point 
correlators \mehta
\eqn\rHo{\rho(\l_1,...,\l_n) =  {(N-n)!\over N!}
{\det} _{n\times n}  K(\l_i, \l_j) }
where 
\eqn\kerN{K(\l , \l ') =
\left\langle \psi^{*\az}(\l)\psi^{\za}(\l')\right\rangle. }
The two-point function  \kerN\ is called spectral kernel
and is expressed in terms of the orthogonal functions
\ortbaZ\  as
\eqn\spckob{K(\l , \l ')= \sum_{n=1}^N {f_n(\l) f_n(\l')\over (f_n, 
f_n)}.}

Another useful representation of $\rho(\l_1, ... , \l_n)$ is
 as the correlators of the Cartan current $H$. 
 One 
finds, using the operator product 
expansion
 for  the currents $H$ and $E_+$, 
\eqn\cleQ{   H(z)  e^{Q_+}|0\rangle = 
\Big\langle \int_{-\infty }^{\infty} { d\l\over z-\l}  E_+(\l)
e^{Q_+}\Big| 0\Big\rangle . }
 This relation generates a set of Ward identities the  simplest of 
 which is
 \eqn\DSVv{ \langle H(z) \rangle = \int _{-\infty}^{\infty}
  {d\l\over z-\l} \langle  E_+(\l) \rangle. }
  This equation together with   \dnsf\ gives the 
  the representation of the resolvent 
   $W(z)= {1\over N} \langle H(z) \rangle$ as an integral of the 
spectral 
density
 $W(z)=   \int _{-\infty}^{\infty}
  {d\l \rho(\l) \over z-\l} $.
   Therefore all the information about the  spectral
   correlators is also contained in the   discontinuities of the
   correlators of $H(z)$.

 \subsec{Virasoro constraints and Hirota equations}

  We  give here another derivation of  the Virasoro constraints, 
using 
 the  Fock space representation of the partition function. 
 The level-1 $\hat{ u}(2)$ current algebra satisfied by the operators 
\currennts\ 
decouples into $\hat{u}(1) $ and $\hat{su}(2)_1 $ parts.
The corresponding Sugawara
   energy-momentum  tensor also decouples to two noninteracting pieces
\eqn\emTz{ \hf \sum_{a=1,2} :(\p\psi^{*\aza} \psi^\aza -
\psi^{*\aza} \p \psi^\aza): \ \ =\  T(z) + \tilde T(z),}
where
\eqn\tzem{\eqalign{ T(z) &\ =\ {1\over 6}:(2 H^2 +E_+E_- +E_-E_+): 
\quad 
=\ 
  :H^2(z):\ , \cr \tilde T (z)& \ = \  : \tilde H^2(z):\cr}}
and $H $ and $ \tilde H $ are given by \utwo\ and  \uone\ 
respectively.
 Both pieces commute with the operator $e^{Q_+}$, which means that 
theory is conformal invariant. Only the operator $T(z) =\sum_{n}L_n 
z^{-n-2}$  
produces  nontrivial Ward identities. For any $n\ge -1$,
 \eqn\vIr{[L_n, Q_+] = \int_{-\infty}^{\infty} d\lambda [L_n, 
E_+(\l)] =
  \int_{-\infty}^{\infty} d\lambda {d\over d\lambda} 
\Big(\lambda^{n+1} 
E_+(\lambda)\Big).}
  Since our potential diverges as $\lambda\to\pm\infty$, the boundary 
terms of 
  can be neglected, and we have
   \eqn\ViR{\langle N| e^{H[V]} \ T(z) \ e^{Q_+} |0\rangle =\{ {\rm 
regular 
\
function \ at \ } z=0\}.}
On the other hand, by  commuting $  L_n  , \ n\ge -1$ with $e^{H[V]}$ 
we 
find  the nontrivial linear 
differential equations \viraa \ known as Virasoro constraints.

 The partition function 
  is by construction a $\tau$-function of the KP 
hierarchy \kmmm ,  $\CZ_V[t] =\tau_N[t]$, 
where the ``times'' $t_n$ are the coefficients in the expansion of 
the 
potential $V(z)=-\sum_{n=1}^{\infty}t_n z^n.$
It is a particular case of the ``general solution'' of the KP 
hierarchy 
presented in \jimiwa \ and  satisfies the bilinear Hirota equations
$$\oint dz \ \prod_{n } e^{{  (t_n-t'_n)z^n}} \ \bigg\langle \det(z-M)\bigg\rangle_{N, \{ t\}} 
\left\langle
\det {1\over  z-M } \right\rangle_{N', \{ t'\} }=0\ \ \ \ (N'\le N).  
$$
or 
\eqn\hireqn{
\oint_{\infty}  dz \   \Big({\bf V}_+(z)\cdot Z_N[t] 
\Big)\ 
\Big({\bf V}_-(z) \cdot Z_{N'}[t']\Big) 
=0                  \ \ \ \ \ \ \ (N'\le N)      
   }
 where the vertex operators $
{\bf V}_{\pm}(z)= \exp\Big(  \pm \sum_{n=0}^\infty t_n z^n \Big) 
\exp\Big( \mp \ln {1\over z} \ {\partial\over \partial t_0}
\mp 
\sum_{n=1}^\infty 
{ z^{-n}\over n} 
 {\partial\over \partial t_n} \Big)$
are understood as formal series in $t_{n}$ and $\p\over \p t_{n}$.
They produce  insertions of fermionic operators in the scalar product.
Namely, $V_+(z)  \CZ_N[t]= \langle N|\psi^\az_{N+\hf}  e^{-H[V]} 
\psi^{*\az} (z) G|0\rangle $,
 $ V_-(z)  \CZ_N[t]=
 \langle N|\psi^{*\az}_{N+\hf}  e^{-H[V]}\psi^{\az} (z) G |0\rangle$.
 The direct derivation of the Hirota equations from the matrix 
integral 
is given in \Irazl .

The  ``susceptibility'' $u[t]= 2 
{\partial^2\over\partial t_1^2}
\log \CZ_N$ satisfies an infinite hierarchy of differential equations 
with respect to the ``times'' $t_n$ the first of which is the 
Kadomtsev-Petviashvili (KP) equation
\eqn\KPe{
3{\partial^2 u\over\partial t^2_2} + {\partial\over\partial t_1}
\Big[ -4 {\partial u\over\partial t_3} +6 u {\partial u\over\partial 
t_1}
+{\partial^3 u\over\partial t_1^3}\Big]=0.
  }
 One can check directly that it is satisfied 
by the
    asymptotic  expansion around the gaussian point 
 \eqn\asser{
{ \CZ_N[t_n]\over \CZ_N[ -\delta_{n2}]}=  (-t_2)^{-N^2\over 2} \exp 
\Big( 
Nt_0 -{N\over 4} {t_1^2\over t_2}+
{N^2\over 4} {t_3 t_1\over t_2^2} - {N \over 
 8} {t_1^3 t_3\over t_2^3}+...\Big).
 }

   The Hirota equations themselves are not sufficient to determine
  uniquely the partition functions $Z_{N}[t]$.
The missing information is supplied by the 
first two Virasoro constraints  $L_{0, -1} \CZ =0$
describing the reaction of the integral \hemm\ to translations and
rescaling of the variables, 
\eqn\virS{
L_0 = N^2 +\sum_{n=1}^{\infty} n t_n \p_n, \ \ 
L_{-1} =  t_1 N+
\sum_{n=1}^{\infty} (n+1) t_{n+1} \p_{n}.}

  Let us consider the example of the cubic potential 
$V(z) = -t_1 z-t_2 z^2 -t_3 z^3$  \KKN.
 The partition function depends, due to the scaling laws \virS\ only 
on
 a single combination  of the three couplings:
$$ \CZ_N [t_1,t_2,t_3] = e^{- {t_1t_2\over 3t_3} 
+ {2t_2^3\over 27 t_3^2}} \CZ_N [x, 0,- 1), \ \ \ x   =  { t_2^2- 
3t_1t_3 
\over 3 t_3^{4/3}}.$$
and the susceptibility depends on the three couplings as
\eqn\suSS{ u (t_1, t_2, t_3)= 2\p_1^2 \log \CZ_N  = t_3^{-2/3}  f(x).}
The KP equation reduces to an ordinary differential equation
$$
{\p\over \p t_1} \Big(f+2f'x-9ff'- {3\over 2} f'''(x)\Big)=0.$$

\newsec{  Bosonic representation}

In most of the applications associated with the large $N$ limit we 
are 
interested not in the exact expressions in terms of determinants of 
strongly oscillating fermionic 
eigenfunctions (typically they have $N$ nodes), but in the smooth 
effective expressions. 
In this sense it is more advantageous to develop the theory of the
collective excitations of fermions for which the small parameter 
$1/N^{2}$ plays the role of a Planck constant and  whose ground 
state   is the equilibrium state of the Dyson gas.
The free boson representation is very useful to study the large
$N$ limit of the theory.

By  the 2D  bosonization formulas we represent the chiral
fermions as vertex operators
\eqn\feRboZ{\matrix{ &\psi^\aza (z) = :e^{\varphi^\aza (z)}:\cr
 & \psi^{ *\aza} (x) = :e^{-\varphi^\aza (z)}:\cr } \ \ 
\ :\psi^{*\aza }\psi^\aza : = \p \varphi^\aza, \ \ \ \ \ \ \ (a= 1, 
2)}
where 
\eqn\boZmodes{\varphi^\aza (z) = \hat q^\aza  +\hat p^\aza  \ln z - 
\sum 
_{n\ne 0}
 {J^\aza _{n} \over n } z^{-n}}
are holomorphic two-component scalar fields with 
\eqn\comR{ [J^\aza_m,J^\zaz _n] = n\delta_{a, b}\delta_{n+m, 0} , \
\ \ [\hat p^\aza , \hat q^\zaz  ] = \delta_{a,b}}
 do that their  operator product expansion is then $ \varphi^\aza (z) 
\varphi^\zaz (z')  \sim  \delta_{a,b}\ln (z-z').$
The normal ordering   
:$J^\aza_n J^\aza_m$:  is defined as $J^\aza_n J^\aza_m$
for $n\le m$, $J^\aza_m J^\aza_n$ for $n>m$ and
$:\hat q \hat p: \ =\  :\hat p\hat q: = \hat q\hat p$.
The boson Fock space is generated by the oscillators 
with negative frequencies applied to the vacuum vector $|0\rangle$  
such 
that
\eqn\vaccs{   \hat p^a  |0\rangle =1, \quad 
J^a _{n}  |0\rangle =0 \ \ \ \ \  (n>0).}
The left vacuum $\langle 0|$ is similarly defined, with the 
normalization 
$\langle 0|0\rangle =0$, and 
 the state $\langle N|$ is constructed as
\eqn\NnNN{\langle N| = \langle 0|e^{N\hat q^\az }e^{-N\hat q^\za]}.}

 It will be convenient to introduce the fields
\eqn\bozT{\phi  = { \varphi^\az - \varphi^\za  \over \sqrt{2}}
,\ \ \ \ \ 
\tilde \phi  = { \varphi^\az  +  \varphi^\za \over \sqrt{2}},}
associated with the $\hat{su}(2)$ and the $\hat u(1)$ 
algebras. The corresponding currents 
$H={J_1-J_2\over 2}, \tilde H ={J_1+J_2\over 2} , E_\pm = J_\pm$ are 
given 
by
\eqn\sutwoc{ \eqalign{H  &= {1\over \sqrt{2}} \ \p\phi(z)
, \ \ E_{\pm}=
: e^{\pm \sqrt{2}\phi}:\cr
\tilde H &= {1\over \sqrt{2}}\ \p\tilde \phi.\cr}
}
 The Hamiltonian and  the screening operator $Q_+$ are  represented as
 \eqn\hamB{H[V] = \sum_{n\ge 0} t_n H_n = {1\over \sqrt{2}}
 \oint_{\infty} {dz\over 2\pi i}  V(z)   \p \phi(z)
, }
 \eqn\scrB{ Q_+ = \int_{-\infty}^{\infty} d\lambda\ :e^{ 
\sqrt{2}\phi(\l)}:}
and the Virasoro  generators are realized as follows
\eqn\vIIr{T(z) = \sum_n L_n z^{-n-2} = \hf : \p\phi(z)
\p\phi(z):}

   The $U(N)$-invariant  correlation functions of 
in the matrix model are obtained through the
identification 
\eqn\resdeT{\Tr {1\over z-M}=  \sqrt{2}\ \p   \phi_{_+} (z), \quad
\det (z-M) = : e^{\sqrt{2}{  \phi} _{_+}(z)}:}
where $\phi_+(z)$  is the singular at $z=0$ part of
the  operator $\phi(z) $.

 The Virasoro constraints \viraa\ follow from the
representation  of the bosonic field $\phi(z)$ by the 
operator
\eqn\PPhi{ {\bf \Phi} (z) ={1\over \sqrt{2}} \sum _{n >0}
t_n z^{n} + {1\over \sqrt{2}}  t_0 + \sqrt{2}{\p\over \p t_0} \ln z + 
 \sqrt{2} \sum _{n\ge 0}
{z^{-n}\over n} {\p\over \p t_n}  }
acting on the partition function. 
The bosonic representation is  
easily generalized for the 
orthogonal and  symplectic matrix integrals, see for example \AMOS .

 \newsec{Quasi-classics }

\subsec{Ground state }

In the large $N$ limit the   bosonic field $\phi =
  {\varphi^\az-\varphi^\za\over \sqrt{2}}$  develops a large (of 
order $N$)
  vacuum expectation value $\phi_c (z) =  \langle 
 \phi(z)\rangle $  which is a solution of classical  Virasoro 
constraints  
\eqn\conF{ T_c \equiv \hf \p\phi^2_c(z) = \  \left\{ {\rm regular\ 
function 
  \  at} \ z=0\right\}} 
satisfying  the boundary conditions at infinity 
\eqn\bcOn{    H_c(z)  \equiv {\p\phi_c(z)\over \sqrt{2}}  = -\hf V'(z)
+{N\over  z} + \CO\left({1\over z^2}\right). }
Therefore  the classical current $ H_c$ is of the form
\eqn\clpp{ H_c(z)   = - M(z)\ y(z), \ \ \ y(z) =
 \sqrt{\prod_{k=1}^{2p}  (z-a_k)}}
where  $M(z)$ is an entire function of $z$.
 
The classical current has discontinuities along the 
intervals  $[a_{2k-1},a_{2k}]$. 
In the case of a single cut, the conditions \conF\ and \clpp\ are 
sufficient for reconstructing the ground state.
In the case of several cuts 
the situation is more complicated. The
potential $V(z)$ then has $p$ minima along the real axis (we assume 
that it
grows at infinity) so that the spectral density
$\rho(\l) = {1\over 2\pi i N} [H_c(\l+i0) -H_c(\l+i0)]$
 splits into $p$
discontinuous pieces $\rho_k$.
Each partition $N_1,...,N_p$
$(N_1+...+N_p=N)$ giving the distribution of the eigenvalues between
the $p$ minima defines a metastable state.
 The tunneling events leading to the 
 decay of the the metastable 
   state
 are of order $e^{- N }$ which means that 
 in the large $N$ limit the metastable states are actually stable.
The true ground state is determined by   requiring that 
the
tunnelling amplitude between two neighboring vacua vanish,
which yields $p-1$ additional 
conditions fixing the filling numbers $N_{1}, \ldots, N_{p}$.

The classical field $\phi_{c} = {1\over  \sqrt{2}}\langle \varphi^{(1)}-  
\varphi^{(2)} 
\rangle $ 
changes its  sign after circling a  branch point. 
This means  that the  the expectation values of the  two original 
bosonic 
fields $\varphi^\az(z)$ and 
 $\varphi^\za(z)$ 
are given by the two branches of the meromorphic function \clpp .
Therefore one can speak of a {\it single}  field $\vec \varphi(z)
= \{ \varphi^\az(z),\  \varphi^\za(z) \}$
defined on the two-fold branched
covering of the complex plane given by the hyper-elliptic
 Riemann surface  
$ y^2 = \prod_{k=1}^{2p} (z-a_k),$
 the  two sheets of which are sewed 
along the $p$ cuts $[a_{2k-1},a_{2k}]$.
 
  \bigskip
  
 \leftline{ \it A. \ One-cut solution}
 
  \bigskip

The classical field depends on $N$ and  on the potential $V(z)$
directly and through   the positions $a_1 , \ a_2 $ of the two branch 
points.
 A simple but important observation (made first by V. Kazakov) is that
the  derivative with respect to 
$N$ of the resolvent of the random matrix (the expectation value  
$H_c(z) $ in our interpretation) depends only on the positions of the 
branch points.
Indeed, according to the normalization condition 
\eqn\normMC{\oint  H_c(z)\ {dz\over 2\pi i} = N  ,} 
 the derivative
 $\p H_c/\p N$ satisfies
\eqn\normOK{ \oint _{A_1} {\p H_c\over \p N} dz = 1}
and therefore behaves as $1/z$  at $z\to\infty$.
 The only analytic function of the form \clpp\ with   this 
asymptotics is
\eqn\solOK{ {\p H_c(z) \over \p N}=
 {1\over y(z)}= {1\over
\sqrt{(z-a_1)(z-a_2)}}.}

If we make a conformal transformation $z\to \omega$ 
with 
$$ \omega = \int^z {d\l \over y(\l)}$$ 
which maps the twice covered 
 $z$-plane to the cylinder $ -i\pi <\i \omega\le i\pi$,
the branch points are replaces by an orbifold structure.
The rotation  around a branch point acts as 
$$\hat \pi :
\vec \varphi(\omega) \to \vec\varphi(-\omega) =
 \Big(\matrix{0 & 1\cr 1 & 0\cr}\Big)\cdot 
\vec \varphi(\omega) ,$$
or
$$\phi(-\omega) = -  \phi(\omega),\quad 
 \tilde \phi(-\omega) = \tilde   \phi(\omega).
$$
 
 According to \solOK ,  the classical field is related tob $\omega = 
\omega(z) $ by 
\eqn\hhH{ {\p H_c\over \p N} = {d \omega\over d z }.
}

The explicit expression for the classical current is
$$H_c(z) = \hf  \oint_{A_1}{dz\over 2\pi i} 
{y(z)\over y(z')} \ {V'(z')-V'(z) \over z-z'}.
$$
Expanding at $z=\infty$ and using the asymptotics
$H_c(z)\sim  {N/z}+ ...$
one finds the conditions 
 \eqn\strngeq{ \oint_{A_1}{dz\over 2\pi i} 
  { z^k V'(z) \over y(z)}=-2N \delta_{k,1}\quad\quad(k=0,1) }
which allow to determine the positions of the branch points
corresponding to given potential.

\bigskip

\leftline{ \it B. \ Multi-cut solutions}
  
\bigskip 

 The above analysis can be carried out in the general case of
 a classical solution with an arbitrary number of cuts.  
  Assuming that 
the branch points are ordered as $a_1<...<a_{2p}$, we have
$$\int _{a_{2k-1}}^{a_{2k}}\rho_k(\l)  d\l  = {N_k\over N}.$$
or, in terms of the zero modes of the classical current,
\eqn\normMC{\oint _{A_k} H_c(z)\ {dz\over 2\pi i} = N_k \quad\quad 
(k=1, 
..., p)}
where $A_k$ is the contour circling  the cut $[a_{2k-1}, a_{2k}]$.

  In this case there the $p$  normalization conditions \normMC\ which 
determine the filling numbers $N_1+...+N_p=N$ associated with the   
minima 
of the effective potential.
 The derivatives of the current  with respect to the filling numbers 
$N_k$ 
\eqn\abdIf{ {\p H_c(z)\over \p N_j}={d \omega_j(z)\over d z}}
form
a basis of holomorphic abelian differentials $d\omega_k(z)$
associated with the canonical $A$-cycles encircling the   
cuts $[a_{2k-1}, a_{2k}]$ and satisfying
\eqn\abdiF{{1\over 2\pi i}\oint_{A_k} d \omega_j(z) =   \delta_{kj} .}
The period matrix of the hyperelliptic surface is given by the
integrals\foot{The period matrix is generally 
overdetermined. The condition
that the the period matrix describes a Riemann surface
(Shottkey problem) is that the corresponding theta function is a
tau-function of KP (Krichever's conjecture proved by
Shiota and Mulase). } 
\eqn\eeqa{ \tau_{kj} = {1\over 2\pi i} \oint_{B_k} 
d\omega_j\quad\quad (k 
= 1,...,p-1, \ j=1, ..., p)}
where the contour $B_k$ encircles the points $a_{2k}$, $a_{2k+1}$,
 ... $a_{2p-1}$ so that 
\eqn\eeqb{A_k \circ B_j = \delta _{k,j},
\ (k,j = 1,..., p-1).}
The explicit form of the abelian differentials is
\eqn\eeqc{{\p \omega_j(z)\over \p z}=
 \sum_k {[K^{-1}]_{kj}z^{k-1}\over y(z)}
\ \ \ K_{kj} = \oint_{A_k}{dz \over 2\pi i}{z^{j-1}\over y(z)}.}
The functions $\p \omega/\p z$  behave as $1/z$ at infinity
and are completely determined by the positions of the cuts.
The explicit expression for the classical current is
\eqn\clCu{H_c(z) = \hf  \oint_C{d\l\over 2\pi i} \ { y(z)\over y(\l)} 
\ 
{V'(\l)-V'(z) 
\over z-\l}
}
where the contour $C$ encloses all cuts and the point $z$.
Expanding in $1/z$ and using the asymptotics \bcOn \
we find
\eqn\mKa{\eqalign{\oint_C{dz\over 2\pi i} { z^k V'(z)\over y(z)} & 
=-2 N 
\delta_{k,p}\
 \ \ \ \ \ \ (k=0, ..., p)\cr}
 }
while the normalization conditions \normMC\ give, after integrating 
along 
the
 contour $A_k$,
\eqn\mcB{
-\hf  \int_{a_{2k-1}}^{a_{2k}}\Pi _k(\l){V'(\l)\over y(\l)} = N_k \ \ 
\ \ 
(k=1, ..., p)}
where 
\eqn\mcC{ \Pi _k(\l) ={1\over \pi} - \!\!\!\!\!\!\!\!\!\ 
\int_{a_{2k-1}}^{a_{2k}} 
{d \mu \over \mu - \l}y(\mu).}
(It is sufficient to add  only $p-1$ such conditions in order to
determine $a_1,...,a_{2p}$.)

The quasiclassical problem with several cuts has been
considered in \refs{ \LPastur , \BDeo ,  \AAk}.
 The authors of \AAk\ find a unique solution (the true vacuum) 
 because they impose 
conditions equivalent to the requirement that the tunelling 
amplitudes 
between the different classical vacua are zero.

\subsec{Quasi-classical expansion}

The quasiclassical expansion of  the free 
energy $\CF_N = \ln\CZ_N$ is actually the 
 $1/N$ expansion
\eqn\Freetot{ \CF_N[V] = \sum _{g=0}^{\infty} N^{2-2g} \CF^{(g)}[V/N] 
+ \ 
{\rm nonperturbative\ terms}.}
It is also called genus expansion because the term
$\CF^{(g)}$ represents the sum of all connected Feynman diagrams of 
genus 
$g$.

The leading term ($\CO( N^{2})$) is simply the action for
 for the classical 
bosonic  field 
\eqn\cFzero{\eqalign{ N^2\CF^{(0)} &= 
N^2  \int \!\!\!\!\!\!-  d\l d\l ' \rho(\l)\rho(\l')
\ln(\l - \l')
-N \int d\l \rho(\l) V(\l)\cr
&={1\over 4\pi}\int d^2 z \p\phi_c\bar \p\phi_c  
-{1\over \sqrt{2}} \oint {dz \over 2\pi i} V(z) \p\phi_c(z).\cr}
}

The  next  term  $(\CO(1)$)  is produced by the gaussian fluctuations 
around the 
classical solution $\phi_c$. It is most easily calculated by 
considering the Riemann surface  
as the complex plane
with two  {\it twist  operators } associated 
with its branch points. The notion of a twist operator
has been introduced by Al. Zamolodchikov \ashktel\
(see also \Knizh).  The twist operators,  which are conformal 
operators 
with    dimension 1/16,
can be represented as vertex 
operators
\eqn\tWist{\eqalign{  \sigma_+ (z)\quad& =\quad :e^{{1\over 
4}[\varphi^\az(z)- \varphi^\za(z)]}:\quad = \quad:e^{ {1\over 
2\sqrt{2}}\phi(z)}:  \ \ \ \ \ (z=
a_{2k-1}),\cr
 \sigma_- (z)\quad &= \quad:e^{{1\over 4}[\varphi^\za(z)- 
\varphi^\az(z)]}: \quad =\quad :e^{- {1\over 2\sqrt{2}}\phi(z)}: 
\ \ \  (z=a_{2k}).\cr}}
  
The twist operators are not conformal invariant and 
have to be dressed by the modes of the  bosonic field 
in order to obtain translational invariant  invariant {\it star 
operators} \Moor .
  One can construct perturbatively the star operators  
$S(a_{k})$  out of the
modes of the twisted bosonic field (the descendants 
of the twist operator $\sigma_{\pm}(a_{k})$), 
by requiring that the singular terms with their OPE with the
energy-momentum tensor vanish.
Then the free energy can be evaluated as the logarithm of the 
correlation function of the $2p$ star operators. 
  
Let us find the explicit expression  for the    $\CO(1)$ 
term $\CF^{(1)}$. Up to  $1/N^{2}$ corrections, the
star operators read
 $$ S(a_{2k-1}) =[ \mu(a_{2k-1})]^{-{1\over 24} } 
\sigma_{+}(a_{2k-1}) , \ \ 
S(a_{2k }) = [\mu(a_{2k })]^{-{1\over 24}  } \sigma_{-}(a_{2k}) \ \ \ 
\ 
(k=1,\ldots, p)$$
where 
$\mu(a)   \sim  \oint_{a}   dz (z-a)^{ -3/2}
 \p\phi(z)  $  is the coefficient in front of the power 
 $(z-a)^{{-1/2}}$ in the mode expansion of $\p\phi(z)$ at the point $z=a$.  
Using the explicit expression for the correlation function of
the twist operators we find
$$ \CF^{(1)} = \log\langle S(a_1) \ldots S(a_{2p})\rangle
= -{1\over 24} \sum_{k=1}^{2p} 
\ln \mu(a_{k}) +{1\over 8} \log \left({\det}_{kj}\left[ {1\over 
a_{2k-1}- 
a_{2j}}\right] \right). $$
This expression generalizes the one-cut solution found in 
 \ackm\ 
   by solving directly the
  loop equations.

  The evaluation in this way of the next term $\CF^{(2)}$  
  requires more  work, but is still much more simple than the direct
  iteration of the loop equations.
   (The result   of \ackm \  was obtained with the help of  
    Mathematica,    and I 
    reproduced all the coefficients,  except the first one,   using 
    only my pen.)

\newsec{Two-point correlators for one-cut background }

\subsec{ The two-point correlator of the resolvent}
  
The connected correlation function of the resolvent 
$$ W(z,z') = \left\langle \tr {1\over z-M} \tr {1\over 
z'-M}\right\rangle_c  $$
is related to that of 
the
current $H= {1\over \sqrt{2}}\p\phi$ 
 by 
\eqn\twptCor{ \eqalign{W(z,z')
&=\hf  \langle \p\phi_{_+}(z)\p\phi_{_+}(z') \rangle\cr
&= \hf \left(\langle \p \phi(z)\p\phi(z') \rangle 
-   {1\over  (z-z')^{2}}\right).\cr}}
(We remind that $\phi_{+}$ is the singular at $z=0$  part of the 
field $\phi$.)
Therefore, in order to evaluate  $ W(z,z')$,   we need  the   
 expression for the  two-point correlator of the gaussian field 
$\varphi(z) $
defined on the Riemann 
surface with a cut along the interval $[a_1,a_2]\equiv[a,b]$.   
The latter  can be evaluated as  the 4-point function of two current
and two twist operators
\smallskip

\eqn\prOp{\eqalign{\langle \varphi(z)\varphi(z')\rangle_c &=  \langle 
0|\varphi(z)\varphi(z')\sigma_{_+}(a) \sigma_{_-}(b)|0\rangle\cr 
 &= \ln \big(\sqrt{(z-a)(z'-b)} -\sqrt{ (z-b)(z'-a)}\big)\cr}.}
This function has a nontrivial monodromy,  and   the determination  of 
the square roots depends on whether  the two
fields  belong to the same or to different sheets    the Riemann surface.
The correlator can be also determined   from the condition that it 
behaves as $\ln(z-z')$ at short distances and that it has the same 
analytic structure as the background field $\varphi_c(z)$. The sum 
over 
the four terms gives of course the two-point correlator on the plane: 
$$\sum_{a,a'} \langle  \varphi^{{\aza}}(z) 
\varphi^{_{(a')}}(z')\rangle=
 \ln[\zeta (z) -\zeta (z')] , \qquad 
 \zeta(z) = (z-a)(z-b) .$$

 We have to calculate the correlator of the 
 field   $\phi(z) = {\varphi^{(1)}-\varphi^{(2)}\over\sqrt{2}}$ which is antisymmetric with 
respect to 
the 
$\Z_2$ monodromy. Taking the alternated sum over the four branches of \prOp, 
we get 
 \eqn\prOp{\langle \phi(z)\phi(z')\rangle_c = \ln {\sqrt{(z-a)(z'-b)} 
-\sqrt{ (z-b)(z'-a)}\over
\sqrt{(z-a)(z'-b)} +\sqrt{ (z-b)(z'-a)}}.}
 
  Differentiating twice, we obtain for the current correlator
 \eqn\curr{ \langle \p\phi(z)\p\phi(z')\rangle_c ={
{\sqrt{(z-a)(z'-b)}\over  \sqrt{ (z-b)(z'-a)}}+  {\sqrt{(z'-a)(z-b)} 
\over 
\sqrt{ (z'-b)(z-a)}}
\over 2(z-z')^2 }  }


\subsec{The quasiclassical expression for the spectral  kernel }

The quasiclassical evaluation of the spectral kernel can be done as 
well,
but in this case it is not so easy to control the approximation. 
Again we split the bosonic field $\varphi$ into a classical
 and quantum parts and   consider the quantum part as a free field 
 on the Riemann surface.
The   kernel \kerN\  is expressed 
in the free field approximation  as 
a correlation function of vertex operators
\eqn\kernB{K(\l , \mu) =   \ \langle :e^{\varphi^\az(\l)}: \ 
:e^{-
\varphi^\za(\mu)}: 
  \rangle.}
 As in the case of the resolvent, we  first  write the quasiclassical 
 expression for
 the four-point function of two vertex and two twist
operators
\eqn\fpF{\eqalign{\CG(z, z')& = \langle 0| 
 :e^{  \varphi(z)}: \ :e^{ - \varphi(z')}: 
\sigma_{_+}(a)\sigma_{_-} (b) |0\rangle\cr
&= {1\over 2 } {e^ {\phi_c(z)-\phi_c(z')\over\sqrt{2} }\over 
z - 
z'}
\Bigg(\bigg[ {(z -a)(z' - b) \over (z - b)(z' - a)}\bigg]^{1/4}
+ \bigg[ {(z -b)(z' - a) \over (z - a)(z' - b)}\bigg]^{1/4}\Bigg)\cr}}
where we have used that 
$\langle \varphi (z)\rangle = \pm {1\over \sqrt{2}} \phi_c (z) 
$, the sign depending on whether the field is evaluated on the upper 
or the lower sheet of the Riemann surface.
The  overall coefficient is determined by the    behavior at $z\to z'$
where we have to reproduce ${1\over \sqrt{2}}\p \phi_c (z)
 = \pm i \pi \rho(z)$. The branch of $\phi_c(z)$ has to be chosen so 
that
 $\CG(z)$ decays exponentially at infinity.
  
 The kernel $K(\l,  \mu)$ is defined on the real axis and is obtained 
  as  the average over the four values of this function
on both sides of the cut $[a,b]$
\eqn\qqKera{ 
K(\l , \mu) = {1\over 4\pi i} \sum_{\epsilon,\epsilon' = \pm}  
\CG(\l+i\epsilon 0,
 \mu +i\epsilon'  0) .}
 (The average should be taken because of the ambiguity due to 
presence of 
the
 cut; this ambiguity of $\CG$  appears only in the  large $N$ 
 limit.) 
 One finds explicitly, for $ \lambda, \mu \in [a,b]$
 \eqn\qqKer{\eqalign{ K(\l , \mu)& =
 {1\over 2 \pi}
 {  {\sin \left( {\phi_c(\l) 
 - \phi_c(\mu)\over i\sqrt{2}} \right)
\over
\sqrt{  (\l -a)(b-\mu )} -\sqrt{  (b- \l )(\mu - a)}}
+  {\cos\left(  
 {\phi_c(\l) + \phi_c(\mu)\over i\sqrt{2}} \right)
 \over
\sqrt{  (\l -a)(b-\mu  )} +\sqrt{  (b- \l )(\mu - a)}}
\over [(\l -a)(b-\l ) (\mu - a)(b-\mu)]^{1/4}}.\cr}}
This expression coinsides with the one obtained in 
\ref\brzee{E. Br\'ezin and S. Hikami, \np B 422 (515) 1994.} and in
\refs{\bhz , \pzj}
by   solving the  appropriate loop  equations.  

When $\l, \mu \in [a, b]$ and $\lambda -\mu \sim 1/N$, 
one obtains the well known short distance behavior
 \eqn\sdbK{ K(\l , \mu) = {\sin [\pi N (\l -\mu)\rho ({\l +\mu\over 
2}])
\over \pi (\l -\mu)} 
.}
In the case when one of the arguments is outside the eigenvalue 
interval,
the  exponent is real and negative and the spectral kernel decays 
rapidly.

\newsec{ Generalization to chains 
of random matrices}

  The most natural generalization of the CFT
construction is given by the   
 $ADE$  matrix
 models \adem , which were introduced  as   a
   nonperturbative microscopic realization of the 
      rational string theories with $C <1$.
     Each one of these models is associated with a rank $r$    
classical 
simply laced  
     Lie algebra (that is, of type  $A_r, D_r, E_{6,7,8}$) or its 
affine 
extension, and represents a system of $r$ coupled random matrices. 

Here we will discuss only the models of the $A$-series,  for which 
there exists a simple fermionic representation. 
The  model associated with $A_r = su(r+1)$ 
represents a chain of $r$ Hermitian matrices 
$M_a$ of size $N_a\times N_a$ ($a=1,...,r$), interacting by means of
$r-1$ auxiliary gauge-field-like
rectangular complex   matrix variables 
$A_{\tilde a} \ (\tilde a =1, ..., r-1) $ of size $N_{\tilde a 
}\times N_{\tilde a +1}$.
In this way the   $M_a$ and $A _{\tilde a}$ 
are associated respectively with 
the nodes and the links of the Dynkin diagram of $A_r$.
The partition function of the matrix chain is given by
the following integral
\eqn\prtFAr{\CZ_{\vec N}[\vec V] =  
 \int \prod\prod_{a=1}^r dM_a\ e^{-\tr V^a(M_a)} \int \prod_{\tilde 
a=1}^{r-1} 
d A_{\tilde a} dA^{^{\dag}} _{\tilde a} e^{ -\tr  A _{\tilde 
a}A^{^{\dag}}_{\tilde a} M_{\tilde a}-
\tr  A^{^{\dag}}_{\tilde a}A _{\tilde a}M_{\tilde a+1}}.
}
After integrating with respect to the $A$-matrices and
 the angular variables of the $M$-matrices, the 
 partition function reduces to an integral 
with respect of the eigenvalues
$\lambda_{ai}, \ i=1, ..., N_a,$ of the hermitian 
matrices $M_a$
\eqn\prtFAr{\CZ_{\vec N}[\vec V] =  
 \int \prod_{a=1}^r \prod_i d\lambda_{ai}
\ e^{-\sum_i  V^a(\lambda_{ai})} \  \prod_{i<j}
(\lambda_{ai}-\lambda_{aj})^2 \  \prod_{\tilde a=1}^{r-1}
 \prod_{i,j} {1\over  (\lambda_{\tilde a i}+\lambda_{\tilde a+1, j})}.
}
{\it Important remark:} the integral with respect to the 
$A$-matrices exists only if all eigenvalues of
the matrices $M_a$ are positive. This can be achieved by 
an appropriate choice of the potential.
We will thus assume that the integration is restricted 
to the positive real axis $\lambda_a>0$\foot{Another way to achieve 
this is to take  $M_a = B_a B_a^{\dag}$ where $B_a$ are complex 
$N_a\times N_a$ matrices. }.

Let $\CC_+$ be a contour representing  the 
boundary of the half-plane $\r z >0$.
 Using the fact that the eigenvalue integration is restricted to the 
positive real axis, we can write 
the following loop equations for each 
 $a$  
\eqn\ADEeqa{\Big\langle W_a(z)^2 + \oint_{\CC_+}
{dz'\over 2\pi i} {1\over z-z'}W(z')\Big[ \sum_bG^{ab}W_b( -z' ) - 
\p_z V^a(z) \Big]\Big\rangle_{N,t} =0.}

  The representation of this integral as a Fock space expectation 
value is a generalization of the $su(2)$ construction of Sect.3. 
Now we consider $r+1$ fermion fields  $\psi^\aza (z), \ a=1,...,r+1,$ 
whose modes in the expansion \pzpo\ satisfy
the anticommutation relations
 $[\psi_{r }^\aza ,\psi^{*^{(b)}}_{s }]_+=\delta_{rs}\delta_{ab} $, 
so that the bilinears
$J^\aza (z)  = :\psi^{*\aza}(z) \psi^\aza(z):$, 
$J_{a}^+(z)  =
\psi^{*\aza}(z) \psi^{^{(a+1)}}(-z)$ and $ J^{-}_a(z) =
\psi^{*^{^{(a+1)}}}(z) \psi^\aza(-z)$
generate an algebra related to the $u(r+1)$ current algebra\foot{The 
difference originates in the minus sign in the definition of the 
currents $J^{\pm}_a$ corresponding to the simple roots of $u(r+1)$.}. 
As before, the $u(1)$ current
$\tilde J = {1\over r+1} \sum_a J^\aza $ completely decouples so that 
the Cartan currents of $su(r)$
are given by the differences
$J^a(z) =  J^\aza (z) - J^{^{(a+1)}}(-z) $.
 
Let us express the potentials $V^a(z)$ as
differences  $V^a(z) = V^\aza(z) - V^{^{(a+1)}}(-z)$,
  and define the     Hamiltonians
$ H^\aza[ V^\aza] = \oint {dz\over 2\pi i} V^\aza (z)   J^\aza(z)$.
  It is easy to see that the  partition function \prtFAr\
 can be written similarly to \fockrp\ as
\eqn\fOckrp{\eqalign{ \CZ_{N}[\vec V] =
\langle \vec l|   \prod_{a=1}^{r+1}
  e^{H^\aza [ V^\aza]}  
\ \prod_{a=1}^r e^{Q^+_{a }} |0  \rangle},}
where the vacuum state of charge $\vec l = (l^{^{(1)}}, ..., 
l^{^{(r+1)}})$ 
is defined by \mnfio\ with  $l^{^{({a+1} )}} - l^\aza = 2 N_a$,    
and the  "screening operators" are  
given    by $ Q_a^+ = \int_{0}^{\infty}  
 d\l\  [ J^\aza (z) - J^{^{({a+1} )}}(-z)] $.

The bosonic representation is 
obtained according to \feRboZ\ in terms of the
$r+1$ bosonic fields $\varphi^\aza$, $a=1, ..., r+1$.
They are split into $r$ fields associated
with the  with the $su(r)$ and the $u(1)$   parts
\eqn\ConCo{\phi^a (z) = \varphi^\aza(z) -  \varphi^{^{(r+1)}} (-z) \ 
\ (a=1, ..., r), \ \ \ \tilde \phi(z) = {1\over r+1} \sum_a 
\varphi^\aza ( (-)^{a-1} z).}It is convenient to define another set of
  fields   $\phi_a$ by
\eqn\ConCon{\phi^a (z)=  2 \phi_a(z) - \sum_b G^{ab} \phi_b(-z)   }
where  $G^{ab}= (\vec \alpha^a\cdot \vec \alpha^b)$ 
is the adjacency matrix of the
$A_r$ Dynkin diagram. The fields $\phi^a$ and $\phi_a$   are related 
as the  contravariant and covariant 
components of a vector field in the base of the simple 
roots $\vec\alpha^a$ of the $su(r)$. Note however the 
reflection $z\to -z$ in the second term of \ConCon.
  
If we define in a similar way the covariant components
$V_a$ of the potential by 
   \eqn\ConCop{V^a (z)=  2 V_a(z) - \sum_b G^{ab} V_b(-z)}
then  
  \eqn\phreZ{\phi_a (z) = -V_a(z) + W_a(z),
 \ \ W_a(z) = \sum_{i=1}^{N_a} {1\over 
z-\lambda_{ai}}.}
The loop equations \ADEeqa\ 
read, in terms of the contravariant components $\phi^a$
\eqn\loopvCF{ \left\langle \int _{\CC_+} {d z'\over  z-z'} 
[\p\phi^a(z')]^2 \right\rangle =0 \ \ \ (a=1, ..., r).}
The Riemann surface defined by the classical 
solution represents an $(r+1)$-fold covering
of the complex plane which has in the simplest case two 
cuts $[a,b]$ and $[-b, -a]$ on the real axis ($0<a<b)$.
The fields $\varphi^\aza$ represent the values
on the different sheets of a single meromorphic field $\varphi(z)$.  
The quasiclassical expressions for the 
correlation functions and the free energy 
are obtained by introducing 
$2r$ twist operators
\eqn\ttWist{\eqalign{  \sigma_{k}^{ +} [ (-)^k a ]
  &=  :e^{   {1\over 
2 }\phi^{k}[(-)^k a]}:  \ , \  \ 
 \sigma_k^{ -} [(-)^k b ]
  =  :e^{  - {1\over 
2 }\phi^k[(-)^k b]}:   \cr}}
associated with the points $\pm a, \pm b$.
In the limit $a\to 0, b\to\infty $ the $r$ twist operators
located at the points $b$ and $-b$ merge into
a  $\Z_r$ twist operator at the origin
while the rest form a $\Z_r$ twisted boundary condition at infinity. 
In this case the bosonic description is
given by the  $\Z_n$ twisted field considered in refs.
\dvv .

 \newsec{Another  generalization: The hermitian matrix model  as an amplitude
between two  $GL(N)$ 
representations}

 Our CFT representation actually  holds    
 for a more general class of matrix models, containing 
 external matrix sources.
  The source can be introduced by replacing the left vacuum state 
with 
  an  excited state.  
  Consider  the integral
\eqn\heCCm{ 
\CZ_{_{R^\az, R ^\za}}[V]  =  \int dM e^{-\tr V(M)}\chi_{_{R^\az}}(M)
\chi_{ _{R^\za}}(M) }
where $\chi_{R^\az}$ and  $\chi_{R^\za}$ are the characters of two
polynomial representations of the
group  $U(N)$.  
The polynomial representation $R$ is labeled by the
Young tableau
$$Y_R
  = \{l_1\ge l_2\ge ...\ge
l_N\}$$
and the corresponding character is given by
a ratio of two determinants 
\eqn\cheRR{\chi_{_R}(M) = { \det \left(\l_i^{j-1+l_{N-j+1}}\right)
\over \det\left(  \l_i^{j-1}\right)} \quad\quad (i,j=1 ,..., N).}
Therefore the partition function can be written again as
 an integral over the  eigenvalues 
\eqn\hCCm{ \CZ_{_{R^\az, R ^\za}}[V]
= \int   \prod _{i=1}^N    d\lambda_i \ e^{- V(\lambda_i)}
\det (\l_i^{j -1+l^\az_{N-j+1} } )\det (\l^{j-1+ l^\za _{N-j+1 }}).
}
The  $U(N)$  characters  are related to the expectation values 
of the fermions as \jimiwa
\eqn\fermChar{\eqalign{ 
\chi_{_{R^{\aza}}}(M)   
&={\left\langle 0\left| \psi^{  \aza }_{ r_1^\aza}...
\psi^{  \aza }_{ r_N^\aza}\ 
\psi^{*\aza}(\l_N)...\psi^{*\aza}(\l_1)\right|0\right\rangle
\over \prod_{i<j}(\l_i -\l_j)}
\cr} }
 where $r^\aza _k = k-{1\over 2} + l^\aza_{N-k+1}$ ($k=1,...,N$).
 The
 Fock representation of the matrix integral \heCCm\ 
  is an evident generalization of \fockrp
\eqn\tauCH{\eqalign{ \CZ_{R^\az R ^\za}[V] =
\langle R^\az R ^\za| e^{H[V]} e^{Q_+}  |0  \rangle}}
where
\eqn\RoRt{\langle R^\az R ^\za|
= \langle 0 | \prod_{k=1}^N \psi^{ \az }_{ r^\az_k}
\psi^{*  \za }_{ r^\za _k}}
generalizes the vacuum state $|N\rangle$.  
The analog  of the diagonalization formula  \ptffF\   is  
\eqn\GptffF{\CZ_{ _{R^\az, R ^\za}}[V] = {\prod_{n=1}^N h_n\over 
\det P_{R^\az}  \det P_{R^\za}} }
where the matrix $P_R $ labeled by the Young diagram 
 $R= \{l_1\ge l_2 \ge
...\ge l_N\} $, or equivalently, by the sequence of half-integers
$r_k= k-\hf + l_{N-k+1}$, is  made of the coefficients of the 
orthogonal polynomials $P_n(\l) = \sum_{k=1}^n P_{n,k}\l^{n-k}$:
 $\ P_{ij}^\aza = P_{r_i-\hf , j}.$ 
When $l^\aza _{N-i-1}\equiv 0$, the determinant of the triangular 
matrix $P_{R^\aza}$  equals 1 and \GptffF\ reduces to the
standard formula  \ptffF.

The quasiclassical expressions for the  
spectral correlations depend on the nontrivial left vacuum  
only through the classical field
$\phi_{c}$ and  are therefore given again by \curr\ and \qqKer .
The   Virasoro constraints  are however more complicated because 
the   left vacuum state is not
annihilated by the negative  components of the current. 

 \newsec{Concluding remarks}
 
 The CFT representation of the matrix models 
 originates in the $SU(N)$ invariant matrix integration measure 
 and therefore  can be actually constructed  for all matrix models.
  In a system of several or even infinitely many interacting random
  matrices, one can associate with each   matrix
  variable $M_{a}$ a current $J_{a}(z) = \Tr {1\over z-M_{a}}$ 
  and a  vacuum $|0_{a}€\rangle $, which is invariant
  with respect to conformal transformations generated by
   the Sugawara stress-energy tensor $T_{a}(x)  \sim  J^{2}_{a}(z)$.
  In the large $N$ limit   all  other  matrices interacting with 
  $M_{a}$ can be reduced to
  an effective matrix mean field, which will plays the role of the 
  potential $V$ in our analysis of the hermitian matrix model.
   The classical Virasoro constraints \conF\  hold again, but 
   the effective potential can have cuts and poles. The Virasoro
   generators do not have in general a simple 
 representation as linear differential operators as   the
   Virasoro constraints \viraa . The classical 
   background is described by a Riemann surface, which is in general 
   not a hyperelliptic one.
   The Hilbert space 
   associated with the matrix variable $M_{a}$ corresponds to 
   a contour separating the cuts of the resolvent  $\Tr {1\over z-M_{a}}$  
   from the cuts of the effective potential.

   The usefulness of the CFT approach consists in the fact that 
   once the classical solution $\varphi_{a}(z)$ is found, 
   the loop correlators and the spectral kernel are given by
   the universal expressions \curr\ and \qqKer. In this sense 
     observed universality  of the short distance behavior
     of the loop correlators  
  and the spectral kernel is a rather trivial consequence of the 
  conformal invariance.
   It is also possible to construct a universal  loop diagram 
   technique for the $1/N$ corrections, one particular case of which 
   is considered in ref. \hk  .

 \listrefs

\bye